# ASTROMLSKIT: A New Statistical Machine Learning Toolkit: A Platform for Data Analytics in Astronomy

Snehanshu Saha[1], Surbhi Agrawal[2], Manikandan. R[3], Kakoli Bora[4], Swati Routh[5], Anand Narasimhamurthy[6]

[1,2,3,6] Department of Computer Science & Engineering, PESIT South Campus, Bangalore
[4] Department of Information Science & Engineering, PESIT South Campus, Bangalore
[5] Department of Physics, Center for Postgraduate Studies, Jain University, Bangalore

[1]snehanshusaha@pes.edu, [2]surbhiagrawal@pes.edu, [3]vishnumani.2009@gmail.com, [4]k_bora@pes.edu, [5]swati.routh@jainuniversity.ac.in, [6]anandmn@pes.edu

## ABSTRACT

*Astroinformatics* is a new impact area in the world of astronomy, occasionally called the final frontier, where several astrophysicists, statisticians and computer scientists work together to tackle various data intensive astronomical problems. Exponential growth in the data volume and increased complexity of the data augments difficult questions to the existing challenges. Classical problems in Astronomy are compounded by accumulation of astronomical volume of complex data, rendering the task of classification and interpretation incredibly laborious. The presence of noise in the data makes analysis and interpretation even more arduous. Machine learning algorithms and data analytic techniques provide the right platform for the challenges posed by these problems. A diverse range of open problem like star-galaxy separation, detection and classification of exoplanets, classification of supernovae is discussed. The focus of the paper is the applicability and efficacy of various machine learning algorithms like K Nearest Neighbor (KNN), random forest (RF), decision tree (DT), Support Vector Machine (SVM), Naïve Bayes and Linear Discriminant Analysis (LDA) in analysis and inference of the decision theoretic problems in Astronomy. The machine learning algorithms, integrated into ASTROMLSKIT, a toolkit developed in the course of the work, have been used to analyze HabCat data and supernovae data. Accuracy has been found to be appreciably good.

## 1. Introduction

Since time immemorial, origin of life has troubled and fascinated astronomers and astrophysicists. Important and relevant questions were asked and answers have been sought over centuries. How did life begin on Earth? Is there any other planet or moon which is capable of supporting life? In other words, does there exist an exoplanet that could support life in the same way as earth?

These questions deserve answers that would initiate the chain of discovery. The advent of technology has equipped the astronomers and astrophysicists with very powerful telescopes, endowed with high precision and resolution, capable of capturing minute details of the sky. A tiny patch of sky generates tremendous volume of data. The study and analysis of this data devoid of automated filtration and calibration posed a big hurdle. Thus, the statistical study of such data has assumed significant proportions. Naturally, a need to foster collaborative engagement among astronomers, astrostatisticians and computer science engineers was pondered upon and implemented in quick time. Astroinformatics is the outcome of such engagement and continues to play a pivotal role in any data-intensive future investigations. **ASTROMLSKIT** is an integrated platform for such data-driven analysis and computation. A suite of algorithms were built (PS: 3.1 for details) to handle several types of datasets and open problems. The focus of the paper is however the deployment of the toolkit in two specific problems in Astroinformatics.

One such problem is finding habitability of an extra solar (exoplanet) planet. Led by Kepler's Mission, around 1800+ planets have been discovered. A symbolic question is whether these celestial objects are itself planets or not. The detection of such planets is non-trivial and inferring habitability is more complicated. The HabCat has several features to analyze in the process of detection and classification. A typical dataset is derived either from photometry or spectroscopy primarily, which is calibrated & analyzed in the form of light curves. Study of these light curves is a challenging problem with regard to their classification. For example, dip in these light curves represents the presence of an exoplanet but this phenomenon may also be due to the presence of eclipsing binaries, pulsating stars, red giants etc. Other problems like brightness level of light curves, presence of noise etc. hamper the process of classification. Exoplanets can be confirmed only after advanced statistical analysis of the data. The different exoplanet detection methods like radial velocity, astrometry, transits, direct imaging and microlensing poses difficult questions in their own capacities. Detailed modeling of planetary signals to extract information about the orbital and atmospheric properties summons very specific skill sets.

Lately, much effort has been directed towards the study of such planets which revolve in the orbit of their stars in their habitable zones. A planet's atmosphere is the gateway to its identity, including the formation, development and sustainability of life. Numerous features such as planet's composition, habitable zone, atmosphere class, mass, radius, density, orbital period, radial velocity, to name a selected few, are considered.

The paper concentrates on the classification of supernovae as well. Identifying the types of supernova from their spectra, for example type Ia, type Ib, type II etc is an intriguing problem. The supernova itself changes over time. Hence, it may belong to a different type depending on the time of observation. The alluring part is that, we have a small dataset, usually in the order of hundred, contrary to most of the other problems currently being solved by data analytic techniques. Supernovae are important in cosmology because maximum intensities of their explosions could be used as "standard candles". Briefly, it helps astronomers indicate the astronomical distances. The problem is centered on two major challenges in data estimation and analysis: Parameter estimation using complex models and data analysis of noisy, nonstandard data types.

There exists a plethora of other problems such as object classification; star–galaxy separation, galaxy morphology, quasars etc, where data can be interpreted with ease and analyzed using machine learning algorithms. However, the authors have chosen not to focus on these problems and their inherent complications in this article.

Relevant and scholarly literature is available in the public domain. We choose to discuss and review some of these. The paper also unveils new software, ASTROMLSKIT, developed in-house with the primary goal of analyzing data relevant to astrophysics and astronomy (http://pythonmlskit.comeze.com). ASTROMLSKIT is an open ended open source statistical toolkit designed to integrate more than 30 algorithms covering multiple areas in Machine learning (classification/clustering), Regression techniques (linear/Generalized Linear), DOE techniques, Distribution Plots and Data Generation, different visualizations (scatter plots/dendrogram / lag plots andrew curves /correlation plots/radviz/bootstrap/hexbin plots). The paper presents deployment of the toolkit to the solution of the two key problems mentioned above.

The remainder of the paper is organized as follows: Literature Survey documents a few problems which have been resolved using machine learning approaches and have helped astronomers in drawing important conclusions. The next section dwells upon the data intensive aspect of the two problems, HabCat classification and detection and Supernovae classification, albeit briefly. The ASTROMLS toolkit framework follows in the next section. The following section, simulation and experimental setup, documents the results of the algorithms tested on the dataset and comments on the pertinence of the toolkit. This section also presents a novel application of Naïve Bayes, DT and LDA algorithms in habitability classification and highlights improved accuracy the algorithms are able to deliver. The discussion and conclusion section summarizes the essence of the paper and elaborates on future research motivation in this domain. Appendices contain a few screenshots of the toolkit, relevant dataset links and short notes on several machine learning algorithms used in the analysis.

## 2. Literature Survey

Machine learning has proved to be a very powerful tool in the classification of important data sets and extracting necessary information and interesting patterns from large amount of data. Machine learning algorithms are classified as supervised and unsupervised. The desired accuracy level of flux is $10^{-4}$ to $10^{-5}$ [1] in the atmosphere of exoplanets, which is difficult to achieve. An improved version of independent component analysis has been proposed in [1], where using an unsupervised learning approach, the authors filtered the noise due to instrumental systematic and other stellar sources. They have used wavelet filter, to remove noise even in low SNR conditions. This is achieved for HD189733b spectrum obtained through Hubble/NICMOS. Light curves describe important data about exoplanets. In [2], a supervised machine learning approach was used on stellar light curves to ascertain the existence of an exoplanet. This is accomplished by representing light curves as time series data explained by various stellar attributes which is then combined with feature selection to obtain the proper outcome. Next, dynamic time warping algorithm is applied which compares each light curve with a baseline light curve, elucidating the similarity between the two. Several other models such as alternate SMO and multilayer perceptron model have been implemented with the accuracy of 83 and 82.2 percent respectively. In [5], light curve dataset obtained from Kepler telescope was used for classification of stars as potentially harboring exoplanets or not. The preprocessing of the large dataset of Kepler telescope light curves removed the initial noise from the light curves and strong peaks (most likely transiting planets) were identified by calculating standard deviations and means for some threshold values. These thresholds were selected from the percentage change metric. Next, feature extraction was performed to help capture the information regarding consistency of the peaks and transit time, which are otherwise relatively short. PCA over these extracted features was executed as the next step as a measure towards dimensionality reduction. Furthermore, four different learning algorithms – KNN, logistic regression, softmax regression and SVM were applied. Softmax regression produced best result for the training data set. Overall performance was boosted by using k-means clustering and further application of softmax regression and PCA to 85% on the test data. NASA's catalog gives recent information about the planetary data, where certain celestial bodies are considered as Kepler's object of interest (KOI). Analysis and classification of these KOI is performed in [3], via a supervised machine learning approach that automates the categorization of the raw threshold crossing events(TCE) into set of three classes namely planet candidate(PC),astrophysical false positive(AFP) and non-transiting phenomena(NTP), otherwise carried out manually by NASA's Kepler TCE Review (TCERT) team. Random Forest classifier was proposed and the classification function was decided based on the statistical distribution of the attributes of each TCE like SNR, angular offset etc. The labels of training data were obtained by matching ephemeris contained in KOI to TCE catalog. Data imputation was carried out by using sentinel value to fill in missing attribute values. Sensitivity analysis was carried out for the same operations. The precision of 95% for PC, 93% for AFP and 99 % for NTP was observed. Further analysis with different classification algorithms (Naïve Bayes, K-NN) was carried out which proved random forest gives the best results. As mentioned in the introductory part of the paper, supernovae classification is also one of the challenging problems in astronomy. Kernel principal component analysis (KPCA) with 1NN (K=1 for K nearest neighbor) was proposed in [4], in order to perform the supernovae photometry classification. Dimensionality reduction was done using PCA and KNN algorithm was applied thereafter. The study concluded that for a dark energy survey sample, 15% of the original set will be classified with the purity of >=90%. In [7], authors described two classes of methods for drawing sharp statistical inferences about the equation of state from observations of Type Ia Supernovae (SNe). The dark energy pressure and density are expressed in terms of co-moving distance, r, which helped calculate reconstruction equation, w, very important in the description of various cosmological models. First, a technique for testing hypotheses about w was derived. "W" is the equation of state that requires no assumptions about its form and can distinguish among competing theories. The technique is based on combining shape constraints on r, exploiting features of the functions in the null hypothesis, and any desired cosmological assumptions. A framework for nonparametric estimation of w with corresponding assessment of uncertainty was constructed. Given a sequence of parametric models for w of increasing dimension, the authors used the forward operator T(·) to convert it to a sequence of models for r and used the data to select among them. In [8], an automatic classification method was proposed for astronomical catalog with missing data. Bayesian networks (BNs), a probabilistic graphical model that is able to predict missing values in the observed data and dependency relationships between variables is used. To learn a Bayesian network from incomplete data, an iterative algorithm that utilizes sampling methods and expectation maximization to estimate the distributions and probabilistic dependencies of variables from data with missing values, was deployed. The goal was to extrapolate values of missing features from the observed ones. In this work, the authors used Gaussian node inference which is commonly used for continuous data. Each variable is modeled with a Gaussian distribution where its parameters are linear combinations of the parameters of the parent nodes in the Bayesian [10]. A semi-supervised method to classify photometric supernova typing was used. The nonlinear dimensionality reduction was performed on the supernova light curves using diffusion map followed by random forest classification on a spectroscopically confirmed trained set to learn a model that can predict types of each newly observed data. It was observed that despite collecting data on a smaller number of supernovae, deeper magnitude-limited spectroscopic surveys are better for producing training sets. For type Ia supernovae it was observed that there was 44% increase in purity and 30% increase in efficiency.

When redshift is incorporated, it leads to a 5% improvement in Type Ia purity and 13% improvement in Type Ia efficiency. Next, they used K2 algorithm, a greedy search strategy to learn the structure of the BN. Finally, Random Forest (RF) classifier was implemented which produced reasonably accurate results.

### 3. Astronomical Analysis and Machine learning

The rapid growth of the internet, widespread deployment of sensors and various scientific researches has resulted in a great volume of data. Gathering and maintaining huge amount of data from a collection is one challenge and extracting useful information from them is even more demanding. Data analytics is the solution for this problem. According to whatIs.com, data analytics (DA) can be defined as the science of examining raw data with the purpose of drawing conclusions about that information. Data analytics focus on inference, the process of deriving a conclusion based solely on what is already known by the researcher. Machine learning methods help researchers to analyze data in real time. Machine learning is a discipline that constructs and study algorithms to build a model from input data. We can conclude that data analytics and machine learning techniques learn characteristics from data. Data analytics uses machine learning methods to make decision for a system.

The learning algorithm is used to discover and learn knowledge from the data. The type and amount of the dataset will affect the learning and prediction performance. Machine learning algorithms are classified into **supervised** and **unsupervised** methods, also known as predictive and descriptive, respectively.

According to [6], Supervised methods rely on a training set of objects (with both features and labels) for which the target property is known with confidence. The method is trained on this set of objects; the resulting mapping is applied to other objects for which target property is not available. In contrast, unsupervised methods do not label data into classes. Unsupervised algorithms usually require some kind of initial input to one or more of the adjustable parameters and the solution obtained depends on this input.

Telescopes with high accuracy and precision are available. Even if the telescope captures a small portion of the sky, tremendous amount of data is generated because of their high resolution. HabCat contains 17,129 candidate host stars for their planets presently. An exponential rise in the data is observed in recent years. Furthermore, the data is noisy (stellar and due to instrumental semantics). That apart, data sometimes is not very clear and several other aspects make it difficult to perform the classification of the data. We suggest machine learning algorithms to obtain high accuracy in appreciably quick time.

However, there is scope for improvement and new methods may be devised as the propensity of data is assuming intimidating proportions. The ASTROMLS toolkit executed a few algorithms on HabCat database as cited in the earlier section and applied new algorithms to classify the exoplanets. Our dataset contains 68 features (columns) and 1896 rows, describing discovered exoplanets from the habitability catalog.

## 3.1 Our framework: AstroMLS Toolkit

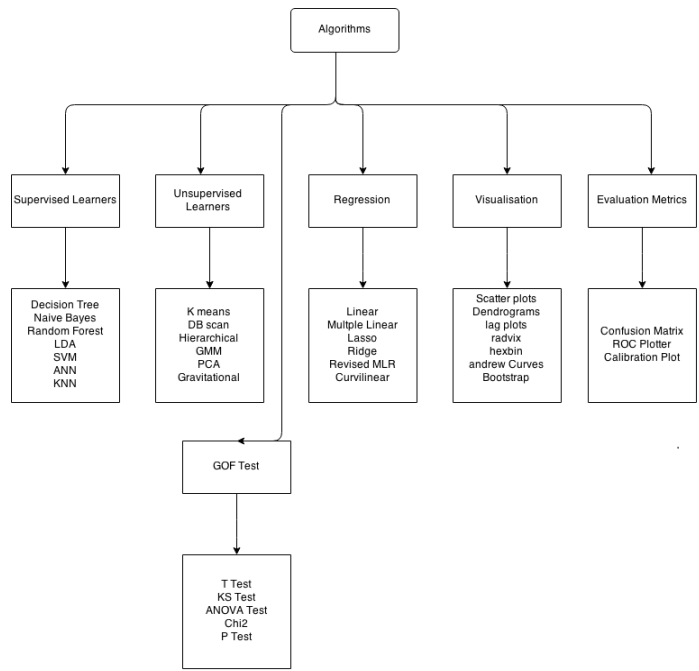

Fig 1.  Algorithmic structure in ASTROMLSKIT

**"ASTROMLSKIT"** is a suite of various machine learning algorithms which are schematically represented in the flow chart above. The toolkit implements algorithms that include Supervised learners, unsupervised learners, Regression learners along with a basic statistical test bed including, but are not limited to, KS test, T test, R-t test , ANOVA  test.

The toolkit is embellished with a visual ease of use interface for the naive user, especially addressing non-IT professionals. This is done via "UI-FORMS", a tool to handle programmers' API's through UI interfaces. Every supervised learning algorithm has an interface where user may upload train and test file with some of the pre-set and standard parameters for that particular algorithm. Let us consider KNN: where "K" value can be set by users while in Naive Bayes' a selection of the kind of Naive Bayes' is done with ease. Split criteria and number of trees can be specified while using the decision tree algorithm. It is typical to choose kernel type while implementing SVM. On the other hand, ANN requires the number of neurons in outer and hidden layers. The interface is minimalist and less confusing, the authors believe.

Un-supervised learning algorithms implemented in the toolkit follow similar UI structure i.e. parameters could be adjusted easily via UI. A representative example is that of "DB scan" which allows customization of clustering parameter, "eps" via UI. In the case of Regression Learners all a user has to do is to set the parameters and specify necessary input. Most of the algorithms have the output written to a file with appropriate labels and descriptors.

The backbone of ASTROMLSKIT is built on python and MVC software architectural pattern, rendering portability to all operating systems and extensibility with new algorithms. The open source nature of the application leaves ample scope for seamless integration from other members of the community.

### 3.2 Visualization Features

The most important part of this toolkit is the visualization paradigm that accompanies data analysis. Visualizations are categorized into "data visualizer" and "evaluation visualizer". The various visualizers, as shown in the flowchart, range from basic scatter plots to advanced radviz plots and hexbin plots. Evaluation visualizers are limited to ROC generators, confusion matrix generators and calibration plots. The data visualizers are in form of UI FORMS, having input for a data file in csv format. Further options on color maps are provided as well.

The visualization tool is fully customizable with features of zoom in/out, saving features etc. The UI forms for Evaluation visualizers are different from that of data visualizers. Each evaluation visualizer has a separate type of input. As an illustration, the case of confusion matrix generators has options where user can either upload the output files of supervised learners to the UI form or a matrix file that's already present. In calibration plotter however, we have to provide a calibration file as an additional input while ROC UI form takes a data file as input. Evaluation visualizers also provide customizable features similar to data visualizers.

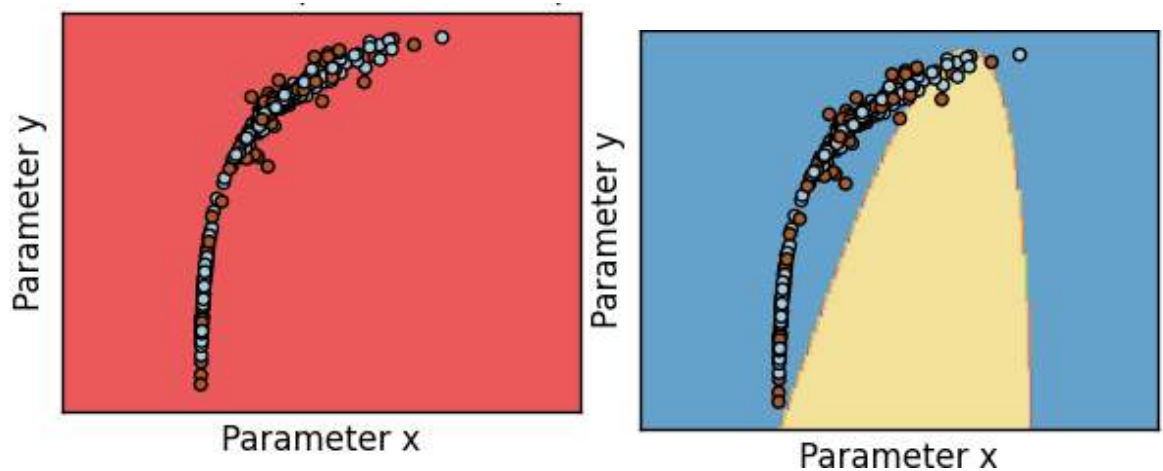

**Fig 2. Visualisation features: Visual representation of a two-class discrimination paradigm(SVM)**

### 3.3 Empirical Study: Data Source

To evaluate the usefulness of the ASTROMLSKIT, two data sets namely HabCat and Supernova were tested with the various supervised algorithms contained in ASTROMLSKIT [17].

The references mentioned in the literature survey are basically investigations on light curves obtained from NASA's catalog, Kepler's object of Interest (KOI) to be precise. In order to test the accuracy of machine learning algorithms we have used HEC (Habitable Exoplanets Catalog), derived from HABCAT, a catalog of nearby habitable systems [12]. This catalog has 17,129 candidate host stars which are potential candidates for harboring life. HABCAT has its source from Hipparcose catalog which contains 118,219 stars. HABCAT was created from the Hipparcose Catalog by examining the information on distances, stellar variability, multiplicity, kinematics and spectral classification for the stars contained therein. The reason for using HabCat is due to an expanded target list for use in the search for Extraterrestrial Intelligence by Project Phoenix of the SETI Institute. HEC data consists of 68 features and 1896 exoplanets (at the time of writing of this paper). The reason of selecting this database is that it combines measures and modeled parameters from various sources. Hence it provides a good metric for visualization and statistical analysis [13]. Such statistical machine learning approaches have not been accomplished on this dataset, to the best of our knowledge, eliciting good reasons to explore and exploit accuracies of different machine learning algorithms.

The HabCat data source possesses 13 categorical features and 55 continuous features. Further, the data suffers from missing values which was tackled using mean for continuous values and most frequent occurrence for categorical values. Supernova data source contains 292 rows and 3 feature columns. Feature Hashing was carried out before the algorithms were fed to the datasets.

### 3.4 Experimental Setup

The experimental study was setup to evaluate performance of our proposed tool and demonstrate the novel approaches for classification of HabCat data.

1. The data sets mentioned above are tested on 6 major classification algorithms namely Naïve Bayes, Decision tree, LDA, KNN, Random Forest and SVM [Appendix B] respectively.

2. Naïve Bayes evaluates the classification labels based on class conditional probabilities with class apriori probabilities, class count, mean and variance set to default values.

3. Decision tree builds a tree based structure by using a split criterion namely GINI, with measure of split being selected as best split and no max-depth and min-depth were specified while Random Forest is an ensemble of decision trees with estimator value set up to 10 trees, the remaining parameters having been set to the same as the decision tree.

4. SVM, a binary classifier was used with a penalty parameter C of the error term, initialized to default 1.0 while the kernel used was RBF [14] and the gamma parameter (kernel coefficient) was assigned to 0.0 and coef0 of the kernel was set to 0.0 as well.

5. The k-nearest neighbor classifier was used with the k value being set to 3 while the weights are assigned uniform values and the algorithm was set to auto.

6. The parameters setup for LDA classifier was implemented by the decomposition strategy similar to SVD [15, 16]. No shrinkage metric was specified and no class prior probabilities were assigned.

## 4. Results and Analysis

A ten-fold cross validation procedure was carried out to make the best use of data, that is, the entire data was divided into ten bins in which one of the bins was considered as test-bin while the remaining 9 bins were taken as training data. We observe the following results and conclude that the outcome of the experiment is encouraging, considering the complex nature of the data.

**Table 1. Results of classification of HabCat data**

| Algorithm | Accuracy (%) |
|---|---|
| Naïve Bayes | 98.7 |
| Decision Tree | 98.61 |
| LDA | 93.23 |
| KNN | 97.84 |
| Random Forest | 98.7 |
| SVM | 97.84 |

**Table 2. Results of supernova classification**

| Algorithm | Accuracy (%) |
|---|---|
| Naïve Bayes | 98.86 |
| Decision Tree | 98.86 |
| LDA | 65.90 |
| KNN | 96.59 |
| Random Forest | 97.72 |
| SVM | 65.90 |

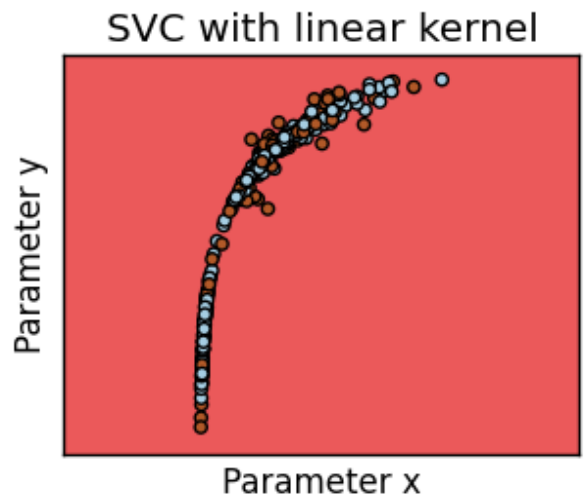

**Fig 3. SVM results on Supernova Data: The above shown figure represents a scatter plot of the results of supernova dataset when applied to SVM. PCA was applied to reduce dimensionality of data. The horizontal axis, parameter x and vertical axis, parameter y refer to values x and y respectively after reducing the supernova data set via PCA. The resulting classes post-classification are represented by different colors. As seen from the above figure there is only one class which is colored by red background and all the data points in reduced dataset are classified into the same class.**

### 4.1 Classification accuracy

Tables 1 and 2 display the results of the two data sets accomplished via 6 major classification algorithms in ASTROMLSKIT.

Performance analysis of the algorithms on the HabCat data is as follows.

Generally all 6 classification algorithms perform fairly well imparting accuracy of >90%. The best accuracy was delivered by Naïve Bayes' (98.7%) and Random Forest (98.7%). This is most likely because of presence of large number of discrete features on the data set, since the Random Forest and Naïve Bayes' can naturally deal with a mix of numerical and categorical attributes. The Decision tree yielded an accuracy of 98.61%, proximal to Random Forest and Naïve Bayes'. Both KNN and SVM produced an accuracy of 97.84% and the LDA provided an accuracy of 93.23%.

Supernova classification analysis offers remarkable discrimination among the algorithmic performance of the six methods deployed.

1. Naïve Bayes and Decision Tree top the accuracy table with accuracy of 98.86%.

2. Random Forest ranks 2 with accuracy of 97.72% and KNN occupies 3$^{rd}$ position with 96.59% accuracy.

3. The dramatic change was observed in the case of SVM, which occupied the last position with LDA with an accuracy of 65.9%. The geometric boundary constraints inhibit the performance of the two classifiers. As shown in Fig 3, where points are occupied under red area which is an indicator of one of the classes. The difference may be observed in Fig 2 when there are multiple classes in the classified data.

Overall, we can conclude Naïve Bayes', Decision Tree and Random Forest perform exceptionally well with both datasets, while KNN acts as average case. SVM and LDA suffer from a dip in accuracy for supernova data.

## 5. Conclusion

In this paper we have presented a review of a new and efficient statistical toolkit relevant to astroinformatics, specifically the habitability problem and supernova classification. A brief survey of various open problems in astroinformatics and applicability of machine learning towards the solution is discussed. The novelty of our work lies in the dataset, HEC (HabCat phlcatalog), that was hitherto not investigated in the existing literature. The most important difference between NASA's catalog and HabCat is that the former makes data available for only those planets which are Kepler's Object of interest whereas the latter contains data for all discovered planets, KOI or not, confirmed or unconfirmed. NASA's catalog for exoplanets has around 25 features whereas HEC has 68 features, including but are not limited to planet's Mass ,radius ,orbital period, planet type, flux, density, distance from star, habitable zone, Earth similarity Index(ESI), habitable class, composition class, eccentricity etc.

Also we unveiled a computational platform, "ASTROMLSKIT" to solve problems in astroinformatics using machine learning. We have computed accuracy of various machine learning algorithms on the HabCat Dataset and Supernova data. Random Forest ranks best with highest accuracy closely followed by Naïve Bayes.

Exoplanets are discovered in regular intervals, if not on a daily basis. A huge task of categorizing those manually may be translated into a simple automated system using this work. A crawler, simple enough to design, may target the major databases, NASA et.al. , and append the Catalog with discovered but non-categorized exoplanets. The suite of machine learning algorithms could then perform the task of classification, as demonstrated earlier with reasonably acceptable accuracy. A significant portion of time could thus be saved. A continuation of the present work would be directed towards achieving a sustainable and automated discrimination system for efficient and accurate analysis of HabCat.

# Appendix A:

## A1. Screenshot of ASTROMLSKIT

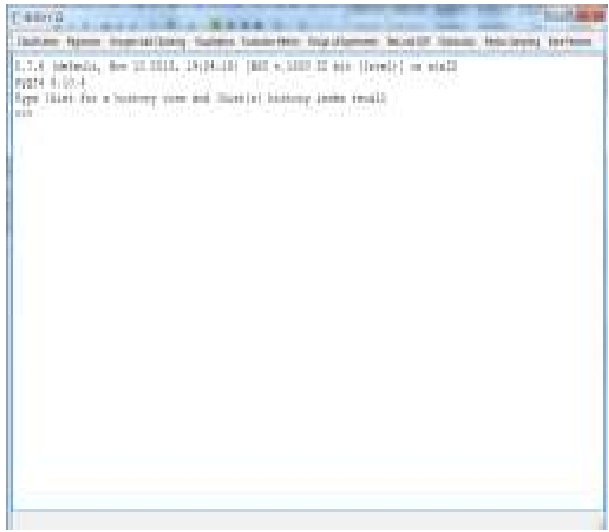

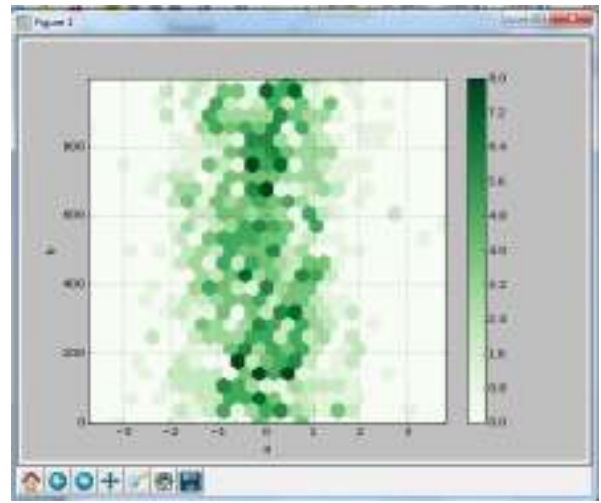

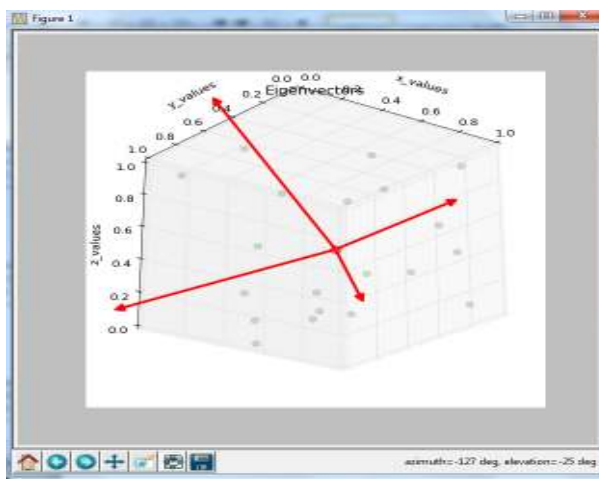

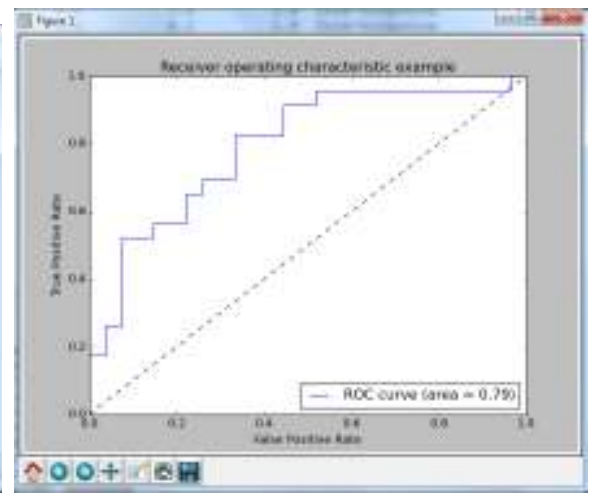

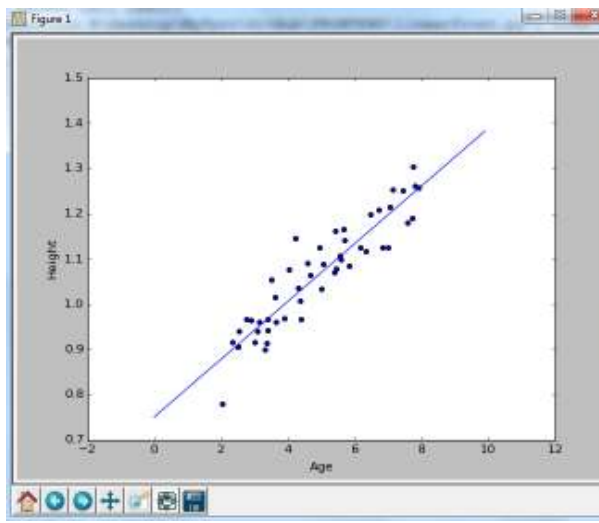

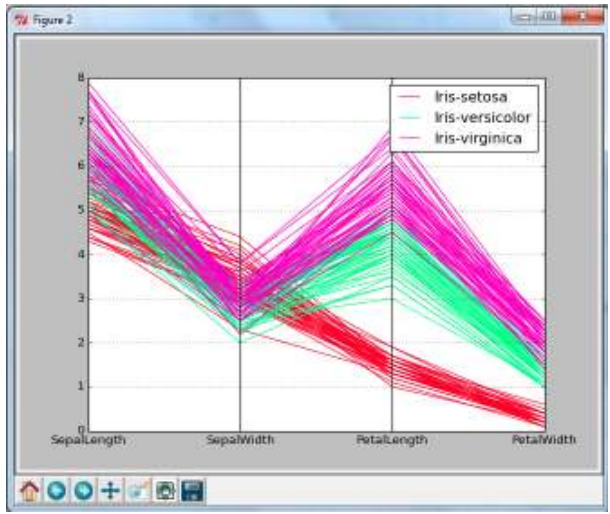
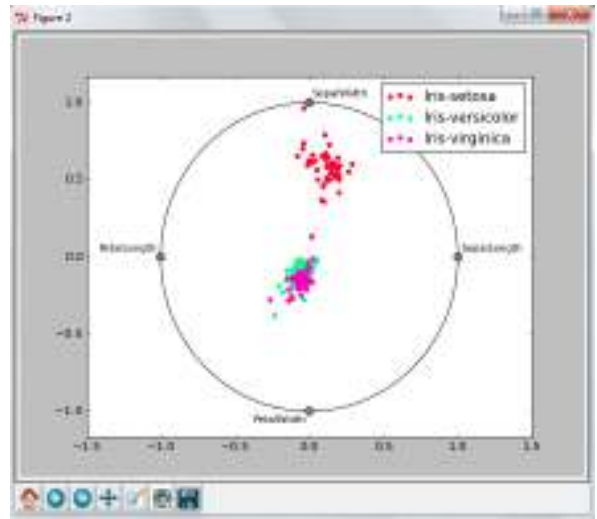
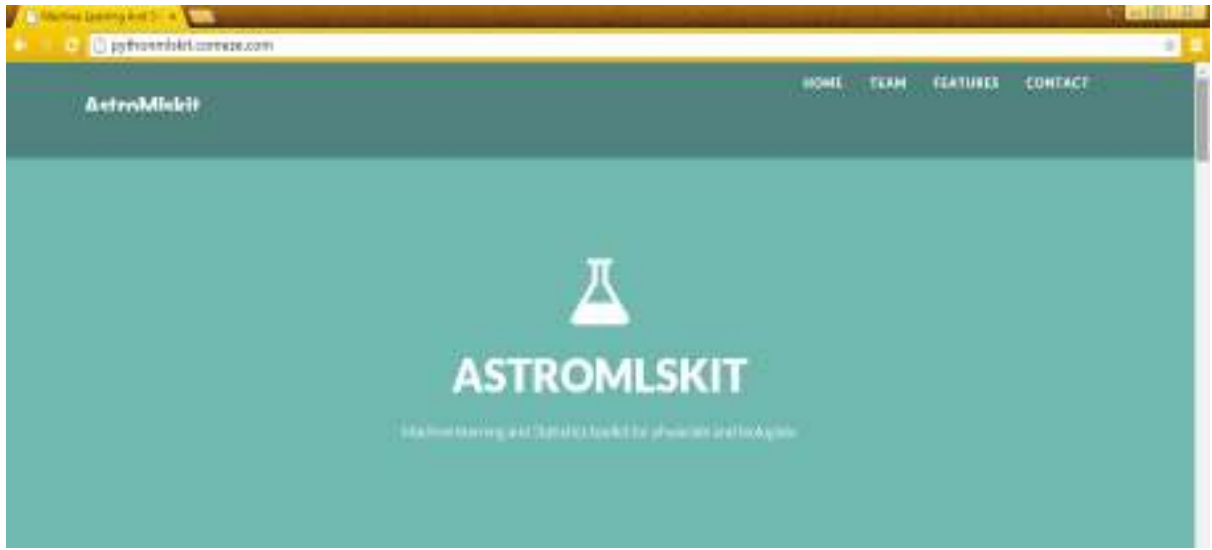

The toolkit is available in http://www.pythonmlskit.comeze.com

**Other websites referred for Data sets( as accessed on 20/04/2015) are listed below.**

1.   http://dark.dark-cosmology.dk/~tamarad/SN/

2.   http://phl.upr.edu/projects/habcat

# Appendix B:

We present short notes on the Machine learning algorithms used in the paper: Naïve Bayes, Linear Discriminant Analysis (LDA), Decision Tree (DT), Random Forest, Support Vector Machine (SVM) and K-nearest neighbor (KNN).

**Linear Discriminant Analysis (LDA) :** The basic LDA classifier attempts to find a linear boundary that best separates the data. This yields the optimal Bayes' classification (i.e. under the rule of assigning the class having highest a posteriori probability) under the assumption that covariance is same for all classes. In our implementation we used a enhanced version of LDA (often called Regularized Discriminant Analysis). This involves Eigen-decomposition of the sample covariance matrices and transformation of the data and class centroid. Finally the classification is performed using the nearest centroid in the transformed space also taking into account prior probabilities.

**Naive Bayes':** Naive Bayes' classifier is based on Bayes theorem. It can perform the classification of arbitrary number of independent variables and is generally used when data is high-dimensional. Data to be classified can be either categorical or numerical. A small amount of training-data is sufficient to estimate necessary parameters. The method assumes independent distribution for attributes and thus estimates $P(X|Y_i) = P(X_1|Y_i) * P(X_2|Y_i) * \ldots P(X_n|Y_i)$. Although this assumption is often violated in practice (hence the name naïve), Naive Bayes' often performs well. It is computationally fast and space efficient.

**Decision Tree:** A decision tree classifier is a machine learning approach which constructs a tree that can be used for classification or regression. Each of the nodes are based on a feature (attribute) of the data set, the first node is called as root node, which can be any important feature and hence considered as best predictor. Every other node of the tree is then split into child nodes based on certain splitting criteria or decision rule, which identify the allegiance of the particular object (data) to the feature class. The leaf nodes represent classification. Typically an impurity measure is defined for each node and the criterion for splitting a node is based on increase in purity of child nodes as compared to the parent node i.e. splits that produce child nodes which have significantly less impurity as compared to the parent node are favoured. The Gini index and entropy are two popular impurity measures. Entropy is used to interpret as a descriptor of information gain from that node.  One significant advantage of decision tree is that both categorical and numerical data can be handled, a disadvantage is that decision trees tend to overfit the training data.

**Random Forest:** Random forest is an ensemble of various decision trees. Each tree enunciates a classification and decision is taken based upon mean prediction on them (regression) or majority voting (classification).  When a new object from the data set needs to be classified, data is kept down at each of the trees. Classification implies a tree voting for that class. Random forest works efficiently with large datasets. It gives accurate results even in the cases of missing data. The training algorithm for random forests applies the general technique of bootstrap aggregating, or bagging, to tree learners. Given a training set $X = x_1, \ldots, x_n$ with responses $Y = y_1, \ldots, y_n$, bagging selects a random sample of the training set with replacement iteratively and fits trees to these samples:

For b = 1, …, B: Sample, with replacement, n training examples from X, Y; call these Xb, Yb.
Next, we train a decision or regression tree fb on Xb, Yb.
Post-training, predictions for unseen samples x' can be made by averaging the predictions from all the individual regression trees on x': or by considering the majority votes in the case of decision trees.

**Support Vector Machine:** SVM classifiers are candidate classifiers for binary class discrimination. The basic formulation is designed for the linear classification problem; the algorithm yields an optimal hyperplane i.e. one that maintains largest minimum distance from the training data, defined as the margin. It can also perform non-linear classification via the use of kernels, which involves the computation of inner products of all pairs of data in the feature space; this implicitly transforms the data into a different space where a separating hyperplane can be found. One advantage of the SVM is that the optimization problem is convex. The result may not be transparent always which is a drawback of this method.

**K-Nearest Neighbor:** KNN is an instance-based classifier that compares new incoming instance with the data already stored in memory. Using a suitable distance or similarity function, KNN relates new problem instances to the existing ones in the memory. K neighbors are located and majority vote outcome decides the classification. Occasionally, the high degree of local sensitivity makes the method susceptible to noise in the training data. If k = 1, then the object is assigned to the class of that single nearest neighbor. A shortcoming of the k-NN algorithm is its sensitivity to the local structure of the data. k-Nearest Neighbors algorithm (or k-NN for short) is a non-parametric method used for classification and regression.